\RequirePackage{fix-cm}
\documentclass[smallextended]{svjour3}  
\smartqed 
\usepackage{graphicx}
\usepackage{hyperref}
\journalname{"AI \& Society"}
\usepackage{graphicx}
\usepackage[table,xcdraw]{xcolor}
\usepackage{tikz}
\usetikzlibrary{matrix,arrows,fit,shapes,decorations.markings,decorations.text,arrows.meta}
\usepackage{longtable}
\usepackage{graphicx}
\usepackage{booktabs}

\begin{document}

\title{{Guiding the Way: A Comprehensive Examination of AI Guidelines in Global Media}

\thanks{This project was partially funded by the University of Amsterdam’s \textbf{RPA Human(e) AI} and by the European Union’s Horizon 2020 research and innovation programs No 951911 (\textbf{AI4Media}).
}}

\author{Mathias-Felipe de-Lima-Santos\and Wang Ngai Yeung\and Tomás Dodds}

\authorrunning{de-Lima-Santos \and Yeung\and  Dodds} 

\institute{
    Mathias-Felipe de-Lima-Santos \at
    Macquarie University, 25 Wally's Walk, Macquarie Park NSW 2109, Sydney, Australia\\
    University of Amsterdam, Science Park 904, 1098 XH Amsterdam, Netherlands\\
    Federal University of São Paulo (Unifesp), Avenida Cesare Mansueto Giulio Lattes, 1201, Parque Tecnológico - Eugênio de Melo,  12247014 São José dos Campos, Brazil\\
   \email{m.f.delimasantos@uva.nl, mathias.felipe@mq.edu.au}    
   \and
    Wang Ngai Yeung\at
    University of Oxford, 1 St Giles', Oxford OX1 3JS, United Kingdom     
    \at Leiden University, Nonnensteeg 1-3, 2311 VJ Leiden, Netherlands\\
    \email{justin.yeung@oii.ox.ac.uk} \\  
    \and
   Tomás Dodds \at
    Leiden University, Reuvensplaats 3 - 4, 2311 BE Leiden, Netherlands \\  
   \email{t.dodds.rojas@hum.leidenuniv.nl} 
}

\date{PREPRINT. LATEST UPDATE: MAY 7 2024}

\maketitle

\begin{abstract}

With the increasing adoption of artificial intelligence (AI) technologies in the news industry, media organizations have begun publishing guidelines that aim to promote the responsible, ethical, and unbiased implementation of AI-based technologies. These guidelines are expected to serve journalists and media workers by establishing best practices and a framework that helps them navigate ever-evolving AI tools. Drawing on institutional theory and digital inequality concepts, this study analyzes 37 AI guidelines for media purposes in 17 countries. Our analysis reveals key thematic areas, such as transparency, accountability, fairness, privacy, and the preservation of journalistic values. Results highlight shared principles and best practices that emerge from these guidelines, including the importance of human oversight, explainability of AI systems, disclosure of automated content, and protection of user data. However, the geographical distribution of these guidelines, highlighting the dominance of Western nations, particularly North America and Europe, can further ongoing concerns about power asymmetries in AI adoption and consequently isomorphism outside these regions. Our results may serve as a resource for news organizations, policymakers, and stakeholders looking to navigate the complex AI development toward creating a more inclusive and equitable digital future for the media industry worldwide. 

\keywords{artificial intelligence\and generative AI\and journalism\and media\and guidelines\and code of principles}

\end{abstract}

\section{Introduction}
Media organizations are avowedly in a continuing transformation process \cite{Deuze2018BeyondJournalism}. The disruption is now characterized by an accelerated pace of innovation in AI-driven solutions, with businesses and developers leveraging these emerging technologies capabilities to create smarter and more efficient systems. In the media industry, the widespread use of artificial intelligence (AI) has recently tethered journalists to generative systems and synthetic technologies that can, for instance, automate content creation, curate personalized news feeds, and analyze vast amounts of data in real time \cite{Pavlik2023CollaboratingEducation}.

The emergence of ChatGPT marked an important milestone in the technology industry potentially triggering a wave of disruption across various fields and applications. Only two months after its launch in late November, the OpenAI’s application crossed 100 million monthly active users, surpassing the initial growth rates of TikTok and Instagram \cite{Hu2023ChatGPTNote}. Positive views suggest that AI, particularly generative AI, can potentially disrupt the media industry due to its capability to generate coherent and contextually relevant text in natural language \cite{Nishal2023EnvisioningMedia}. This versatility makes it a valuable tool for a wide range of tasks within the news value chain, including content creation, virtual assistants, and language translation. Nonetheless, the actual impact of AI in journalism has shown a somewhat less pronounced scale than what was previously expected \cite{Simon2022UneasyAutonomy}. However, the rapid evolution and integration of these tools into newsrooms necessitate ongoing vigilance and assessment.

ChatGPT might be grabbing headlines for its language prowess, but it is merely the first glimpse into a future with the growing presence of generative AI. Tools like Bard (now Gemini) and art-generating applications like DALL-E and Midjourney are just a few examples making waves for their potential to revolutionize the media industry. While AI has the potential to significantly transform industries, its impact may be limited by various factors such as resistance, knowledge, regulatory constraints, ethical considerations, and technological limitations \cite{Makwambeni2023BetweenAfrica,Munoriyarwa2023ArtificialNewsrooms}. Text-based generative AI is already proving its worth in research, brainstorming drafts, and planning. While current generative AI excels at creating images based on simple descriptions, and there are even attempts at video generation, these applications are limited by the heavy data processing required. However, these models are on continuous advancements.

As this field progresses, creating near-photorealistic images and videos of practically anything is likely on the horizon, potentially blurring the lines between AI-generated content and the real world. AI models are also demonstrating continuously improving abilities in composing music and mimicking human speech, further expanding the creative possibilities. Another example is AutoGPT, which helps to create their own prompts, unlocking the ability to tackle even more intricate tasks without  human input. Thus, Generative AI's potential spans various domains beyond media, including healthcare, finance, transportation, and education.

While AI disruption can open up new opportunities for automation and enhanced user experiences, it also raises important questions about ethical considerations, data privacy, and its responsible use, prompting ongoing discussions of regulations and algorithmic fairness and privacy in the media ecosystem \cite{Lin2022TheDemocracy}. For example, AI-produced misinformation and deep fakes can undermine public trust, while automated content creation lacks empathy to understand human behavior’s cultural nuances or ethical reasonings \cite{Pelau2021WhatIndustry}. Similarly, AI-powered news recommender systems may bolster existing social inequalities, hindering social groups such as women or people of color while preventing them from achieving equal representation \cite{Stoica2018AlgorithmicNetworks}. The rise of AI systems in journalism also poses a significant threat regarding job displacement \cite{Munoriyarwa2023ArtificialNewsrooms}. Media organizations may opt to use AI tools for data analysis and routine article writing to save on costs and time \cite{Aissani2023ArtificialConcerns}. These cost-saving strategies could translate into a decline in specialized journalism, such as investigative reporting, where human intuition, creativity, and ethical judgment are irreplaceable factors to do the job \cite{Cancela2021BetweenReporting}.

By deploying stringent editorial guidelines encompassing rigorous ethical framework and professional standards, news managers recognize the benefits and risks that these AI tools could have for journalistic work, addressing the challenges of such technologies in the same way they have handled other technological developments in the past \cite{Oppegaard2023WhereData}. The role of guidelines in journalism, as Foreman et al. \cite{Foreman2022TheAge} put it, unlike absolute rules, is to help practitioners make decisions in a rational way. In other words, they are favoring a relational rather than paramountcy approach.

Codes of ethics and guidelines are central in journalism, as they act as indispensable instruments of accountability for news work. Their significance is underscored by the fact that virtually every major professional news organization has embraced a set of codes and subjected them to continuous adaptation and refinement over time. While overarching principles might be shared among these organizations’ guidelines, individual news entities may also craft their distinct ethical codes, thereby delineating explicit guidelines that demarcate ethical expectations for their workforce \cite{Whitehouse2010NewsgatheringAge}. As the boundaries of journalism expand, these guidelines are becoming more thoughtful and deliberate about using emerging technologies in the news value chain, particularly AI.

This study embarks on an analysis of openly available guidelines formulated by organizations extending across 17 countries, aiming to understand media entities’ responses to the increasing prevalence of AI within newsrooms, particularly generative AI (GAI). Drawing on institutional theory and digital divide and inequality concepts on media and political systems, this work seeks to illuminate the strategies that these organizations employ to safeguard fundamental journalistic tenets, including transparency, accountability, fairness, and privacy.  

In the rapidly evolving landscape of the media industry, the digital era has brought both opportunities and challenges. Digital inequality, characterized by disparities in access to and proficient use of digital technologies, has become a critical issue shaping the media landscape. This creates a digital divide that not only impacts practitioners' ability to access information but also influences the production and dissemination of news \cite{Jamil2023EvolvingPakistan,Mabweazara2013NormativeEra}. At the same time, the digital age has ushered in new norms and guidelines for journalism, reshaping the way news is gathered, verified, and disseminated in digital settings. These new norms play a crucial role in shaping the credibility and trustworthiness of news sources, and emerging leaders are the ones who define them. Moreover, media self-regulation, as expressed through guidelines, encompasses ethical norms internally determined by newsroom professionals rather than externally imposed by governmental bodies \cite{Ruggiero2022TheInsecurity}.

By examining how journalism becomes increasingly institutionalized in the digital era and how digital divides shape journalistic practices, we can gain a deeper understanding of the pervasive influence of AI technologies into the broader dynamics of the media industry. The "AI-fication" age involves more than just applying AI as a technological veneer; it is about infusing intelligence throughout an organization's DNA, which they may not necessarily be ready for. This underscores the urgent need to address challenges and opportunities, particularly the digital inequalities that impact access to and utilization of these tools. Additionally, this convergence highlights the importance of ensuring that AI-driven innovations in journalism are accessible, inclusive, and aligned with ethical and professional standards.

Accordingly, our contention in this study is that examining the ethical guidelines forged by news organizations grants insight into the decision-making dynamics within newsrooms. Thus, this study proposes three research questions (RQs):

    \textbf{RQ1: }How has the media industry addressed the increasing presence of AI, especially generative AI, in newsrooms regarding the inclusion of responsible and ethical factors like fairness, privacy, reliability, transparency, and accountability in their guidelines?

    \textbf{RQ2:} How do media organizations employ AI guidelines to institutionalize values and address the challenges posed by these technologies in journalism, reflecting their commitment to preserving journalistic principles while adapting to the evolving media landscape?

    \textbf{RQ3: }How does the dominance of Western, Educated, Industrialized, Rich, and Democratic (WEIRD) countries in AI development and utilization impact Global South nations in embracing and institutionalizing AI-related principles?

This endeavor offers a glimpse into the sense of responsibility journalists and managers hold while shedding light on how media entities strive to bolster credibility and reliability by navigating the challenges posed by generative AI today.

Furthermore, this study makes a significant contribution by providing a comprehensive analysis of the strategies employed by media organizations to address these challenges through their AI guidelines, examining specific approaches taken by diverse media entities across different regions and cultural contexts. Thus, this research underscores the importance of understanding the global trends in AI utilization. This study also addresses the need for nuanced, local responses and adaptations necessary to ensure ethical and responsible journalism in a rapidly changing media environment applied to different levels of digital development. Additionally, this research contributes to the broader discourse on digital inequality and the impact of AI by exploring how non-WEIRD nations grapple with these challenges and the implications for their journalistic practices.

\section{Theoretical Framework}

\subsection{Digital Inequality and Divide: Navigating the AI Landscape in the Media Industry}

As the digital revolution unfolds, a stark disparity persists between the Global North and the Global South, often referred to as the “digital divide” \cite{Mabweazara2021TowardsJournalists}. This divide represents a significant challenge for countries in the Global South, where access to digital resources and opportunities remains limited for many. The digital divide manifests across various dimensions through unequal access to new media technologies. For example, the most apparent aspect of this divide is limited internet and access to digital devices \cite{Mabweazara2013NormativeEra}. Similarly, inadequate digital infrastructure, including limited broadband coverage, outdated telecommunications networks, and aging technological devices, hinders efforts to bridge this divide. In numerous countries in the Global South, which broadly encompasses regions such as Africa, Latin America and the Caribbean, Asia, and Oceania (excluding rich nations like Israel, Japan, and South Korea), both rural areas and marginalized communities frequently face challenges related to unreliable internet connectivity and the affordability of smartphones or computers. These issues, however, are not limited to rural areas and also affect metropolitan regions. 

While it is true that some countries in the Global South have made significant strides in technology development, such as Brazil, which has a thriving tech industry and widespread internet access in urban areas, it is also crucial to recognize that many countries across this area still face challenges in terms of access to technology and digital infrastructure. Our intent is not to paint the entire Global South with a broad brush but rather to highlight the disparities and challenges that exist within and between countries in the region and the Global North.

When applied to the media industry, this can translate into organizations that lack the technological expertise and innovative capacity to integrate digital solutions into their business models \cite{Holder-Webb2012TheEthics}, limiting them to basic use or dependence on third-party companies \cite{de-Lima-Santos2021Out-of-the-boxAustralia}. Smaller media outlets may also struggle due to their reliance on limited resources, which can compromise their ability to embrace digital transformation. Lastly, many individuals in the Global South lack the necessary digital literacy skills to navigate the online world effectively. This includes skills related to information searching, online communication, and digital problem-solving. This digital literacy gap is also evident among news professionals who may not be tech-savvy and require additional support to engage with technology-driven solutions \cite{Mabweazara2013NormativeEra}.

As the world continues its rapid advancement into the digital age, the transformative power of AI is undeniable. AI technologies promise to reshape industries, economies, and societies, offering unprecedented opportunities for progress and development. However, this transformative potential is not distributed equally across the globe. While the Western, Educated, Industrialized, Rich, and Democratic (WEIRD) countries have been at the forefront of AI development and utilization \cite{Starke2022FairnessLiterature}, the Global South, comprising non-WEIRD nations, faces profound challenges in bridging the digital inequality and divide that AI exacerbates.

Infrastructure development is a crucial driver of AI deployment \cite{Ferrer-Conill2023DatafiedRegulations}. Western nations invested significantly in building robust digital infrastructures, while many Global South countries still struggle to provide reliable internet access to their populations. This disparity affects the ability to access AI-driven solutions and restricts participation in the global digital economy \cite{Brodie2023DataPlanet}.

Technologies’ transformative potential relies heavily on human capital and expertise in related fields \cite{Holder-Webb2012TheEthics}. However, the education and skill development necessary for AI innovation and application remain a substantial challenge in the Global South. Unequal access to quality education, especially in STEM (Science, Technology, Engineering, and Mathematics) fields, further widens the digital divide. Similarly, limited resources and outdated curricula often hinder the development of AI-related skills among the youth. This perpetuates the dependence on foreign expertise, limiting the region’s ability to harness AI for its development \cite{De-Lima-Santos2023GoogleEast}.

On another note, AI has the potential to drive economic growth and create new job opportunities. However, it can also exacerbate economic inequalities in the Global South. Adopting AI technologies often demands substantial investments, which larger, more established businesses can afford, leaving smaller organizations at a disadvantage \cite{Gondwe2023CHATGPTAI}. This disparity is particularly noticeable in the news industry, where mainstream media is capitalizing on AI solutions and implementing systems for its workforce. In contrast, smaller media companies depend on tech platform funding to experiment with specific AI solutions \cite{De-Lima-Santos2023GoogleEast}.

AI technologies also raise profound ethical questions regarding data privacy and sovereignty. In the Global South, data protection and governance issues are complex. Many non-WEIRD countries lack the legal frameworks and resources to regulate data collection, storage, and use effectively. This leaves their populations vulnerable to exploitation by multinational corporations (e.g., Big Techs) and governments from more technologically advanced regions \cite{Birch2022BigTech,Brodie2023DataPlanet,Hendrikse2021TheEverything}.

The ethical considerations extend to AI biases and discrimination. AI systems trained primarily on data from WEIRD countries may not perform well in diverse non-WEIRD settings, leading to biased or blatantly untrue outcomes \cite{Starke2022FairnessLiterature}. Addressing these risks requires local AI research and development to ensure that AI technologies are adapted to the specific sociocultural contexts of non-WEIRD nations \cite{Gondwe2023CHATGPTAI}.

WEIRD countries, being at the forefront of technological development, have well-established institutional frameworks that not only facilitate access to information and communication technologies (ICTs) but also govern their responsible use. These countries have regulatory bodies, legal protections, and comprehensive digital literacy programs in place \cite{Krarup2023EuropeanMaking}. In these countries, where institutions are well-established and digital infrastructure is advanced, the media industry has been at the forefront of AI adoption \cite{de-Lima-Santos2021AIndustry}. In contrast, many countries in the Global South are still grappling with the institutionalization of digital norms and regulations. The lack of well-defined institutional structures and policies can exacerbate the digital inequalities, making it harder for these marginalized communities to access and utilize effectively, particularly in the media industry \cite{Vos2019TheorizingTheory}.

Thus, institutional theory, when juxtaposed with the context of the digital divide, also sheds light on the contrast between WEIRD countries and the Global South. Institutional theory delves into the fundamental and enduring elements of social structure, examining how structures, such as systems, regulations, norms, and habitual practices, solidify their position as authoritative standards for social conduct \cite{Scott2013InstitutionsIdentities}. Below, we discuss how institutional theory not only informs the development of AI in the media industry but can also underscore the significance of institutionalization in shaping AI adoption patterns in media.

\subsection{Unpacking Journalistic Norms and Guidelines: Insights into Institutionalization of Journalism in the Digital Age}

Journalistic norms and values serve as the ethical backbone of newsrooms, guiding reporters and editors in their pursuit of accurate and fair reporting. These principles, including objectivity, impartiality, and transparency, uphold the public’s trust in journalism \cite{Hanitzsch20192.Roadmap}. In newsrooms, institutionalization processes reinforce these norms, making them an integral part of the journalistic culture \cite{Singer2015OutBounds}. This observation aligns with Institutional Theory, which delves into how organizations and institutions develop and maintain structures, norms, and practices that persist over time \cite{Lowrey2017TheEcology}.

In journalism, norms, and values theoretically represent the fundamental principles and values that form the ethical foundation of the profession. These are reinforced through codes of ethics, which journalists should follow and seek assistance when needed. Similarly, guidelines are practical tools and instructions designed to aid journalists in applying these norms to their daily work. This seeks to define, include, exclude, and sometimes blur the parameters of what constitutes a “good” practice of journalism \cite{Lamont2002TheSciences}. While norms and codes of ethics provide a moral compass, guidelines offer specific steps and recommendations for ensuring effective and relevant journalism practices \cite{Singer2015OutBounds}.

However, these norms and guidelines are malleable and negotiable. These codes of ethics, instructional guidelines, and professional standards serve to instruct and occasionally compel journalists to make choices that are considered rational and suitable within their organizations, but they do not necessarily reflect these practitioners’ values and beliefs. Particularly in the Global South, where professionals are very dependent on the political landscape of their particular nations, there is a need to have certain flexibility \cite{Mutsvairo2022JournalismSouth}.

Considering the dynamic nature of journalistic norms and practices, formalizing codes of ethics and guidelines entail professionalization \cite{Karlsson2023RecodingMedia} and institutionalizes the boundary of journalistic work \cite{2015BoundariesJournalism}. They are established to self-regulate deviations from accepted journalistic behaviors. Functioning as rigid rules and procedures, guidelines ensure journalism adheres to the highest standards, ultimately benefiting the public, and prevent media practitioners from performing undesirable behaviors \cite{vanderWurff2011BETWEENAUDIENCE}. As a result, both play essential roles in upholding the integrity, credibility, and responsible conduct of journalism.

Institutionalism provides valuable theoretical insights into how institutions are shaped by these normative pressures, which can originate from external sources like society or internal factors within the institution itself \cite{Meyer1977InstitutionalizedInstitutio}. These pressures compel institutions to adhere to legitimized, rationalized elements, encompassing standard operating procedures rooted in both old traditions (often guided by laws) and new traditions (influenced by norms, beliefs, and practices).

Organizations strive for legitimacy by emulating other actors in their environments \cite{Deephouse1996DoesLegitimate}. Their behavior responds by integrating “correct” programs and structures from their fields, which are perceived as legitimate by other institutions, reflecting efficiency and indicative of quality output \cite{Lowrey2005CommitmentIsomorphism}. Over time, this incorporation of elements drives institutional isomorphism, influenced by pressures that aim to increase an organization’s likelihood of survival within its environment. Isomorphism, defined as a constraining process that drives one unit in a population to resemble other units facing the same set of environmental conditions, can manifest through three mechanisms: \textit{coercive} (pressures from competitors or cultural expectations), \textit{normative} (progressive professionalization in the field), and \textit{mimetic} practices (emulating paths already tested by preceding institutions and actors) \cite{DiMaggio2000TheFields}.

Institutionalism has achieved recognition in various social sciences, including sociology \cite{DiMaggio2000TheFields}, economics \cite{North1999DealingEnvironment}, and political science \cite{March1998TheOrders}. In the journalism realm, these external pressures materialize through regulative, normative, and cognitive institutions. Media organizations are subject to external pressures and expectations stemming from their societal context. Consequently, these organizations influence and are influenced by various stakeholders, leading to the institutionalization of specific journalistic practices and norms \cite{Laaksonen2022MediatedOrganization}.

Media organizations strive to adhere to similar structures and practices to establish legitimacy within their institutional environment. The ongoing negotiation and adaptation of norms and operational procedures can be likened to Lewin’s  concept of \textit{social equilibrium} \cite{Lewin1947FrontiersDynamics}. While initially developed to understand group dynamics and transformations within evolving groups, it provides valuable insights for examining journalism as a socio-technical institution. Journalism essentially comprises a collective of individuals engaged in processes that often appear to achieve a state of relative stability \cite{Paulussen2016InnovationNewsroom}. This process of isomorphism can result in the standardization of news formats, story conventions, and production processes, even across diverse news organizations \cite{Lelo2022WhenMovement}. Similarly, this concept extends to guidelines and codes of ethics, which become influenced by prevailing societal ideologies or the attitudes of early-adopting organizations \cite{Holder-Webb2012TheEthics}.

Furthermore, this theory helps us understand how changes and innovations occur within journalism. When new practices or technologies emerge, they often face resistance due to institutional pressures to maintain the status quo \cite{Kosterich2022NewsJournalism}. However, these innovations can eventually gain acceptance and become institutionalized if they align with the evolving demands and expectations of the media landscape.

Institutionalization is inherently interactive and is constructed based on past experiences. Therefore, it is evident how practices can be continuously negotiated and transformed over time. In the following section, we delve into how these principles of institutionalized rituals are reflected in AI guidelines within the context of the media industry.

\section{Data and Methods}

\subsection{Data Collection and Sample}

This research analyzes the global landscape of AI guidelines within the media industry. For this reason, no language and regional criteria were set for the sampling procedure. To create a comprehensive dataset of guidelines for our analysis, we employed a combination of methods, including snowball sampling and consultation with experts in the field. Snowball sampling allowed us to identify relevant guidelines by starting with a set of known sources and then expanding our search based on recommendations and references provided by those sources \cite{Biernacki1981SnowballSampling}. Additionally, we utilized search engine tools to identify guidelines publicly available online. This multi-faceted approach ensured we captured a diverse selection of guidelines from the media industry.

Our sample encompasses a diverse array of media companies, including, but not limited to, legacy media, digital news outlets, public service media (PSM), and news agencies. We have exclusively included publicly available guidelines within this sample. While this may not be exhaustive, we maintain confidence that this study offers a comprehensive and nuanced understanding of the integration of AI within media companies across various countries and continents. We have excluded commentaries, opinion pieces, and regulatory texts to ensure that our sample exclusively consists of guidelines and protocols with practical implications for media practitioners. These documents typically follow distinct production processes and have differing effects on journalists. 

Furthermore, to be included in our sample, the documents need not be an entirely independent text on the guidance of AI in a certain newsroom. In other words, AI-related protocols embedded in pre-existing codes of ethics are also included in our study. In the end, we collected a total of 41 media-related AI guidelines and policies, and after applying our inclusion-exclusion criteria, 37 of them were included in our final analysis. These guidelines come from organizations located in 17 different countries (see Table \ref{tab:Table1}).

\begin{table}[]
\resizebox{\textwidth}{!}{%
\begin{tabular}{@{}clllr@{}}
\toprule
\multicolumn{1}{l}{\textbf{Continent}} &
  \textbf{Country} &
  \textbf{Organization} &
  \textbf{Type} &
  \multicolumn{1}{l}{\textbf{Release date}} \\ \midrule
Europe          & Germany         & Bayerischer Rundfunk & PSM                                                                 & 30-11-2020 \\
Europe          & Germany         & DPA                  & Agency                                                              & 03-04-2023 \\
Europe & The Netherlands & ANP                  & Agency                                                              & 01-03-2023 \\
Europe          & The Netherlands & MediaHuis            & \begin{tabular}[c]{@{}l@{}}Legacy/\\ Traditional Media\end{tabular} & 14-06-2023 \\
Europe          & The Netherlands & Volkskrant           & \begin{tabular}[c]{@{}l@{}}Legacy/\\ Traditional Media\end{tabular} & 01-05-2023 \\
Europe          & France          & Le Parisien          & \begin{tabular}[c]{@{}l@{}}Legacy/\\ Traditional Media\end{tabular} & 24-05-2023 \\
Europe          & Switzerland     & Heidi.News           & Digital Media                                                       & 08-04-2023 \\
Europe          & United Kingdom  & BBC                  & PSM                                                                 & 01-10-2019 \\
Europe          & United Kingdom  & OFCOM                & Regulator                                                           & 03-04-2023 \\
Europe          & United Kingdom  & The Guardian         & \begin{tabular}[c]{@{}l@{}}Legacy/\\ Traditional Media\end{tabular} & 16-06-2023 \\
Europe          & United Kingdom  & The Guardian         & \begin{tabular}[c]{@{}l@{}}Legacy/\\ Traditional Media\end{tabular} & 28-07-2023 \\
Europe          & Czech           & Czech News Agency    & Agency                                                              & 24-04-2023 \\
Europe          & Sweden          & Aftonbladet          & \begin{tabular}[c]{@{}l@{}}Legacy/\\ Traditional Media\end{tabular} & 05-04-2023 \\
Europe          & Austria         & APA                  & Agency                                                              & 01-03-2022 \\
Europe &
  Belgium &
  Raad voor de Journalistiek (RVDJ) &
  \begin{tabular}[c]{@{}l@{}}NGO/\\ Coalition\end{tabular} &
  21-03-2023 \\
Europe          & Spain           & Prodigioso Volcán    & Innovation company                                                  & 01-11-2020 \\
Europe          & Switzerland     & Ringier              & Innovation company                                                  & 30-05-2023 \\
Europe          & Norway          & Verdens Gang         & \begin{tabular}[c]{@{}l@{}}Legacy/\\ Traditional Media\end{tabular} & 12-04-2023 \\
Europe          & United Kingdom  & BBC                  & PSM                                                                 & 05-10-2023 \\
Europe          & Germany         & DW                   & PSM                                                                 & 19-09-2023 \\
North America   & USA             & Adobe                & Multimedia Software Company                                         & 08-05-2023 \\
North America   & USA             & WIRED                & Digital Media                                                       & 17-03-2023 \\
North America   & USA             & News Media Alliance  & \begin{tabular}[c]{@{}l@{}}NGO/\\ Coalition\end{tabular}            & 20-04-2023 \\
North America   & USA             & Partnership on AI    & \begin{tabular}[c]{@{}l@{}}NGO/\\ Coalition\end{tabular}            & 19-02-2023 \\
North America &
  USA &
  Radio Television Digital News Association (RTDNA) &
  \begin{tabular}[c]{@{}l@{}}NGO/\\ Coalition\end{tabular} &
  17-05-2023 \\
North America   & USA             & CNET                 & Digital Media                                                       & 06-06-2023 \\
North America   & USA             & Business Insider     & Digital Media                                                       & 13-04-2023 \\
North America   & USA             & AP                   & News agency                                                         & 22-02-2017 \\
North America   & USA             & USA TODAY NETWORK    & Digital Media                                                       & 27-04-2023 \\
North America &
  Canada &
  Globe and Mail &
  \begin{tabular}[c]{@{}l@{}}Legacy/\\ Traditional Media\end{tabular} &
  14-06-2023 \\
North America   & Canada          & Reuters              & Agency                                                              & 16-10-2022 \\
North America   & Canada          & CBC                  & \begin{tabular}[c]{@{}l@{}}Legacy/\\ Traditional Media\end{tabular} & 12-06-2023 \\
South America   & Brazil          & Nucleo               & Digital Media                                                       & 18-05-2023 \\
South America   & Brazil          & Agência Tatu         & Digital Media                                                       & 11-10-2023 \\
Asia            & Philippines     & Rappler              & Digital Media                                                       & 12-09-2023 \\
Asia            & Hong Kong       & Hong Kong Free Press & Digital Media                                                       & 17-10-2023 \\
Oceania         & New Zealand     & Stuff                & \begin{tabular}[c]{@{}l@{}}Legacy/\\ Traditional Media\end{tabular} & 15-06-2023 \\ \bottomrule
\end{tabular}%
}
\caption{Identified AI Guidelines for Media Organizations.
}
\label{tab:Table1}
\end{table}

We acknowledge that our search criteria, which were based on the languages spoken by the authors, may have limited the scope of our sample. We will further explore this limitation in the last section of this paper. As some guidelines are in their original languages, the authors manually translated the Spanish, Portuguese, and Dutch guidelines to English. Conversely, for the languages the authors do not speak, Google Translate was used to translate documents, such as Swedish, French, Norwegian, and German. 

Our final dataset incorporates various metadata, including details like the organization's location, type, title of the guidelines, region, and publication date. We extracted the publication date from the guidelines themselves, and when this information was not available, we utilized WaybackMachine or the last date of update from the webpage's HTML file.

\subsection{Convergent Parallel Design}

The study employs a mixed-method design known as Convergent Parallel Design (CPD), as proposed by Creswell and Clark \cite{Creswell2011ChoosingDesign}. Unlike other sequential or multiphase mixed-method designs, such as the exploratory sequential design, CPD is a straightforward yet powerful single-phase research design that enables concurrent data collection, analysis, and interpretation of quantitative and qualitative data.

In our study, we implemented CPD specifically at the data analysis and initial interpretation stage. We separated computational and qualitative methods during the initial iteration of analyses and interpretations. After the first analysis phase, we compared and integrated both sets of results to identify supporting and contradictory evidence, facilitating a comprehensive understanding of the topics and themes within these AI guidelines.

We utilized topic modeling for computational analysis to uncover latent topics within these documents. Concurrently, for qualitative analysis, we applied thematic analysis to identify, analyze, and report patterns or themes within our dataset, allowing us to examine the content of the guidelines in depth \cite{Braun2006UsingPsychology}. 

\subsection{Computational Approach: Topic Modeling with LDA}

In this study, using computational methods, particularly Latent Dirichlet Allocation (LDA), to model the topics within AI guidelines represents a methodological choice grounded in its generally high interpretability and ability to handle sparse input \cite{Egger2022APosts}. Essentially, topic modeling is a bottom-up approach to topic retrieval.

First and foremost, the sheer volume of AI guidelines available for analysis demands a systematic and scalable approach to highlight their main communities in light of the regulative, normative, and cognitive institutions often resulting in isomorphism. As exemplified by LDA, topic modeling provides an automated means to extract latent themes or topics from a vast corpus of textual data. This allows for comprehensively exploring the diverse range of issues and recommendations within AI guidelines.

Furthermore, using computational methods enhanced the objectivity and reproducibility of the analysis. Computational methods also enabled a data-driven approach to topic retrieval. Instead of imposing predefined categories or themes onto the data, topic modeling allowed topics to emerge organically from the content. This bottom-up approach ensured that the analysis remains grounded in the actual content of the guidelines, capturing both expected and unexpected themes. In addition to these advantages, computational methods offered the potential for quantitative insights, such as the prevalence of specific topics or the co-occurrence of themes within the guidelines. This quantitative dimension complements the qualitative analysis and enriches our understanding of the patterns and trends present in the AI guidelines.

For this, we followed standard preprocessing procedures of topic modeling before running the models. Specifically, we removed all stop words and non-alphabetic characters, lemmatized, and tokenized these words. With the generated topics, we identified that these topics are in close relation and demonstrate a line of reasoning why institutionalization of rules and codes of ethics should be applied to (generative) AI. Although LDA cannot not inform how these topics are manifested in these guidelines, this method can give us a glimpse of the emerging topics observed in these documents. We identified three main topics: (i) the need for editorial values in place, (ii) the need to protect users and (iii) the need for guidelines to ensure ethical and responsible development of AI in newsrooms. Out of the three topics, we further allocate the most relevant topic of each guideline according to the probability of the document landing on a certain topic with a probability threshold of 0.6. One point to note is that as our corpus is rather small and the themes of the documents are highly similar, some documents were not allocated a distinctive topic as they are equally probable to land on other topics. In other words, our assumption is that most of these documents cover the same content to a large extent even if they landed on certain topics as some of these topics essentially contain a considerable number of overlapping words. Further details of this methods can be found in the \href{https://osf.io/hc9vb/?view_only=a0719618ce9a4b41bcdf6e3404c365f5}{Annex 1}. 

\subsection{Qualitative Approach}

In this section, we describe the approach we employed to conduct a qualitative analysis of guidelines. The guidelines analysis is a valuable research approach to get a deeper understanding of the principles and best practices in the journalistic field. To analyze the guidelines, we relied on thematic analysis, which is a widely used qualitative research technique that helps identify, analyze, and report patterns or themes within a dataset, making it particularly suitable for examining guidelines \cite{Braun2006UsingPsychology}.

Our analysis is composed of several steps. Once the guidelines were collected, we organized them into a structured format for analysis. This included documenting publication dates, authors or organizations responsible for the guidelines, and any contextual information that might be relevant to the analysis. Thematic analysis involves a systematic coding process. One researcher independently coded the guidelines to identify recurring themes or patterns. Coding involved both deductive coding, where we applied pre-existing categories based on our research questions, and inductive coding, where we allowed themes to emerge organically from the data \cite{Braun2006UsingPsychology}. To ensure intercoder reliability, the researchers met regularly to discuss and refine the coding scheme.

After coding a substantial portion of the guidelines, we began the process of theme development. This involved reviewing the coded segments and grouping them into broader themes and sub-themes. The researchers engaged in an iterative refinement process, revisiting the data and adjusting the themes as needed to capture the richness of the guideline content \cite{de-Lima-Santos2022ProPublicasStories}.

Once the themes were established, we conducted a deeper analysis of the data within each theme. This involved exploring the nuances and variations within and across guidelines. We paid particular attention to any contradictions or divergent perspectives that emerged, seeking to understand the complexities of the guideline content \cite{Braun2006UsingPsychology}. After analyzing the data within each theme, we interpreted the findings in the context of our research question. This step involved drawing connections between the themes, identifying overarching patterns or trends, and considering the implications of the guideline content for our research objectives.

In the final phase, we documented the results of our thematic analysis in a comprehensive narrative described in our findings section. This included a detailed description of the identified themes, illustrative quotes from the guidelines, and an interpretation of the findings in relation to our research question. Finally, we applied a convergent discussion of the two above-mentioned approaches to provide a clear and coherent narrative that conveyed the insights gained from these guidelines.

\section{Results}

\subsection{A general overview of AI guidelines in media organizations}

\begin{figure}[h!]
\centering
\includegraphics[scale=0.25]{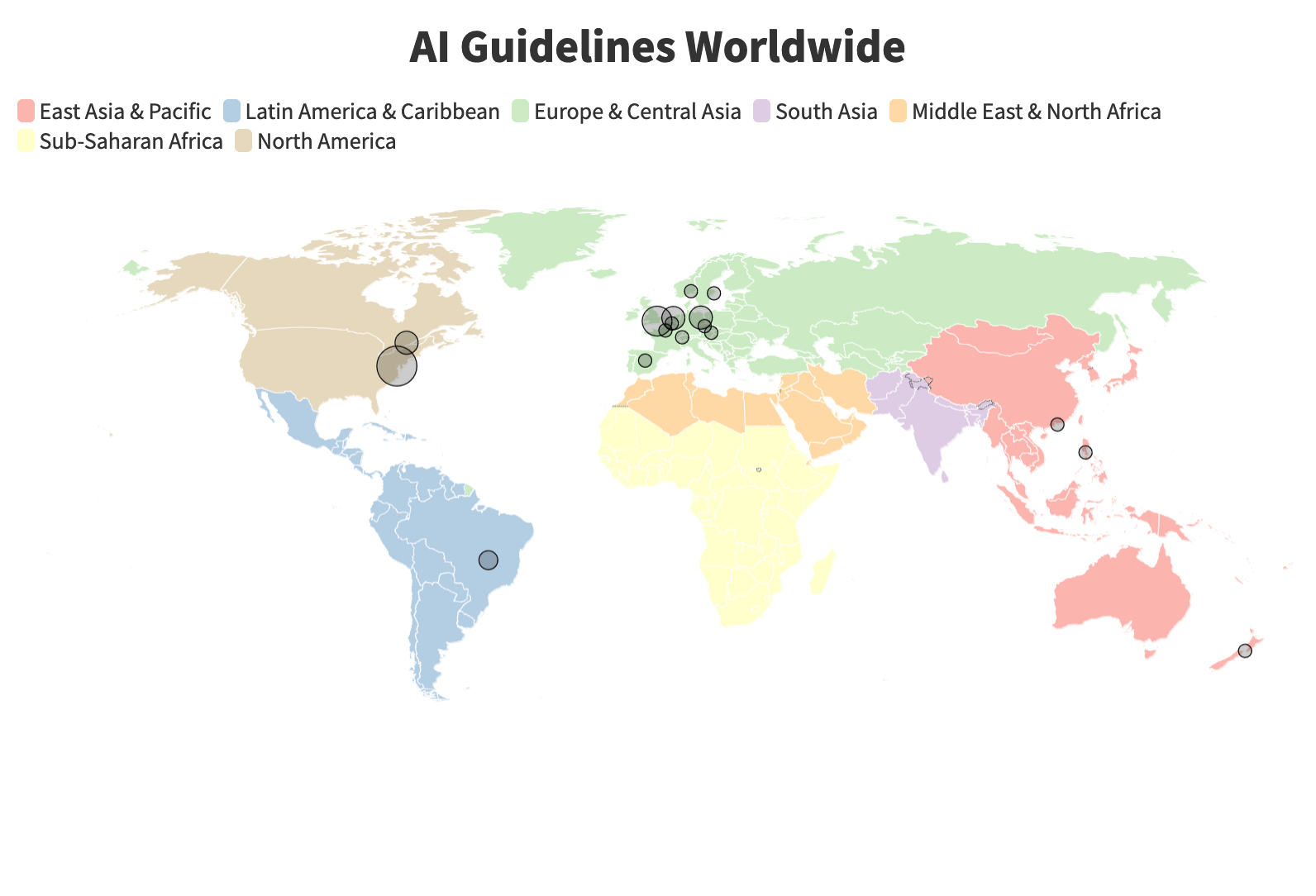}
\caption{ AI Guidelines Identified for Media Organizations Worldwide}
\label{fig:Fig1}
\end{figure}

To provide a comprehensive narrative of the global landscape of AI guideline adoption within media organizations, we illustrated the evolution of these principles by looking at the geographical and temporal patterns. Figure \ref{fig:Fig1} visually encapsulates a significant trend in the landscape of AI guidelines: the predominant presence of media organizations in Western nations, particularly across North America and Europe. This geographical concentration underlines the early leadership of Western countries in formulating and implementing AI guidelines within their media operations. This positions them as authoritative voices, with their perspectives being considered desirable and normative. From a cognitive standpoint, it signifies that their viewpoints are widely understood and accepted \cite{Deephouse1996DoesLegitimate}.

\begin{figure}[h!]
\centering
\includegraphics[scale=0.8]{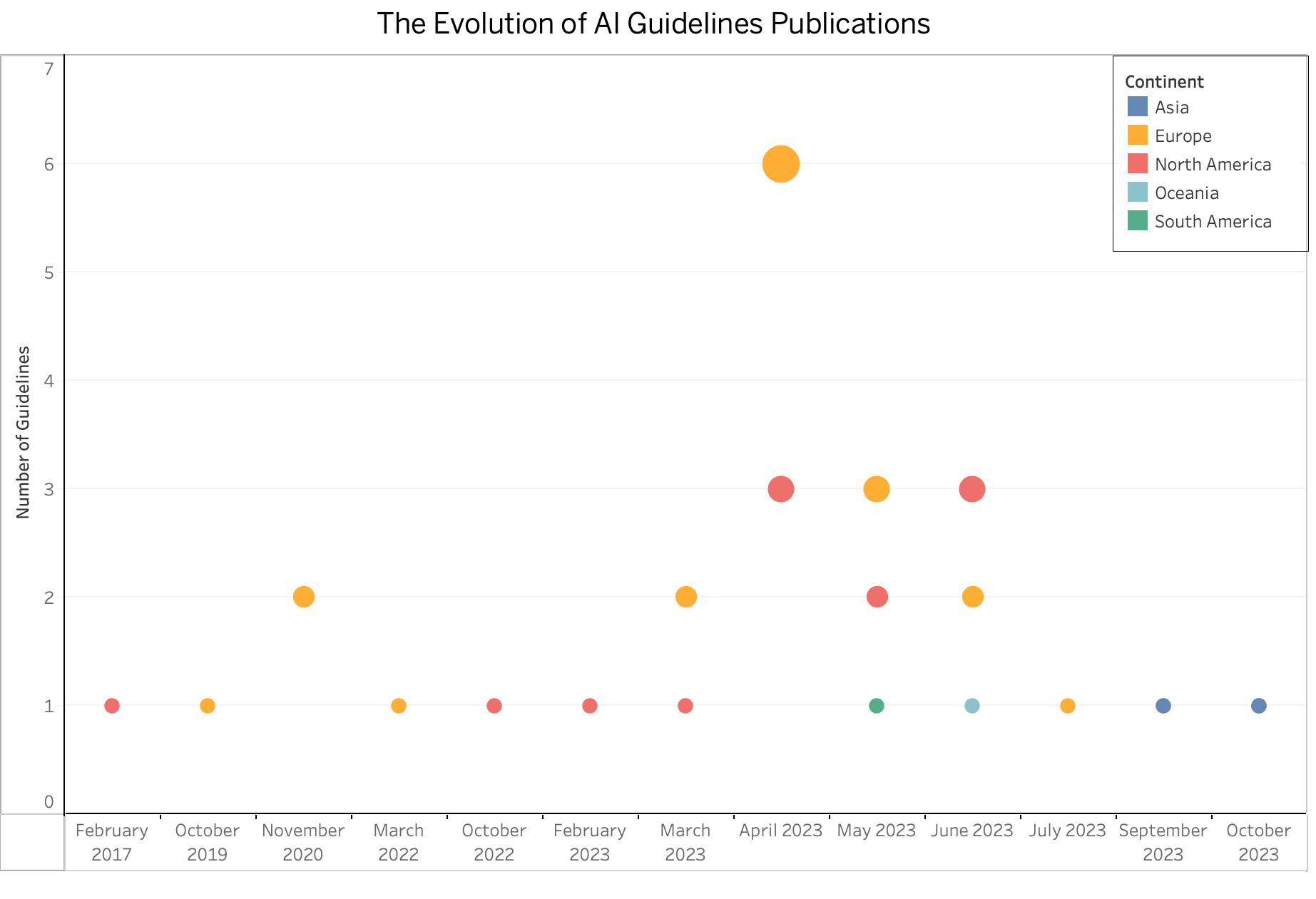}
\caption{Temporal journey of AI guidelines publications. Each dot on the graph represents the number of guidelines published in a specific month in these regions.}
\label{fig:Fig2}
\end{figure}

Figure \ref{fig:Fig2} further underscores this pattern by illustrating the temporal journey of AI guideline publications. It becomes evident that the initial pioneering efforts in deploying AI guidelines were primarily concentrated in North America and Europe. As the chronological progression unfolds in Figure \ref{fig:Fig2}, it is apparent that media organizations outside of these regions have begun to follow suit only more recently. Collectively, these figures tell a compelling story of the geographical bias in AI guideline adoption, primarily favoring Western nations. This bias also reflects the significant influence these early adopters have had in shaping the practices and strategies adopted by media organizations in other parts of the world. In essence, these organizations are playing a crucial role in promoting institutional isomorphism, serving as the anchor point for these other institutions \cite{Suchman1995ManagingApproaches}.

\subsection{Findings Derived from LDA}

We have identified several common themes and strategies that highlight the complex landscape of AI-generated content. These insights are drawn from the perspectives and practices of various organizations, each providing a distinct approach to tackle these challenges. The main topics are described in \ref{tab:Table2}.

\begin{table}[]
\resizebox{\textwidth}{!}{%
\begin{tabular}{@{}cll@{}}
\toprule
\textbf{Topic} &
  \multicolumn{1}{c}{\textbf{Label}} &
  \multicolumn{1}{c}{\textbf{Relevant extracted keywords}} \\ \midrule
1 &
  \begin{tabular}[c]{@{}l@{}}Need for guideline: \\ Responsible and ethical development of generative AI in newsroom\end{tabular} &
  \begin{tabular}[c]{@{}l@{}}System, application, service, guideline, ensure, \\ user, right, risk, development, ethical, standard, \\ develop, quality, training\end{tabular} \\
2 &
  \begin{tabular}[c]{@{}l@{}}Editors-in-the-loop: \\ Ensuring journalistic values in the loop\end{tabular} &
  Journalism, news, fact, newsroom, idea, research \\
3 &
  \begin{tabular}[c]{@{}l@{}}Protection of creators and users: \\ Legality, rights and transparency\end{tabular} &
  \begin{tabular}[c]{@{}l@{}}Audience, public, approach, trust, right, value, risk, \\ responsible\end{tabular} \\ \bottomrule
\end{tabular}%
}
\caption{Identified topics from LDA.}
\label{tab:Table2}
\end{table}

\begin{figure}[h!]
\centering
\includegraphics[scale=0.5]{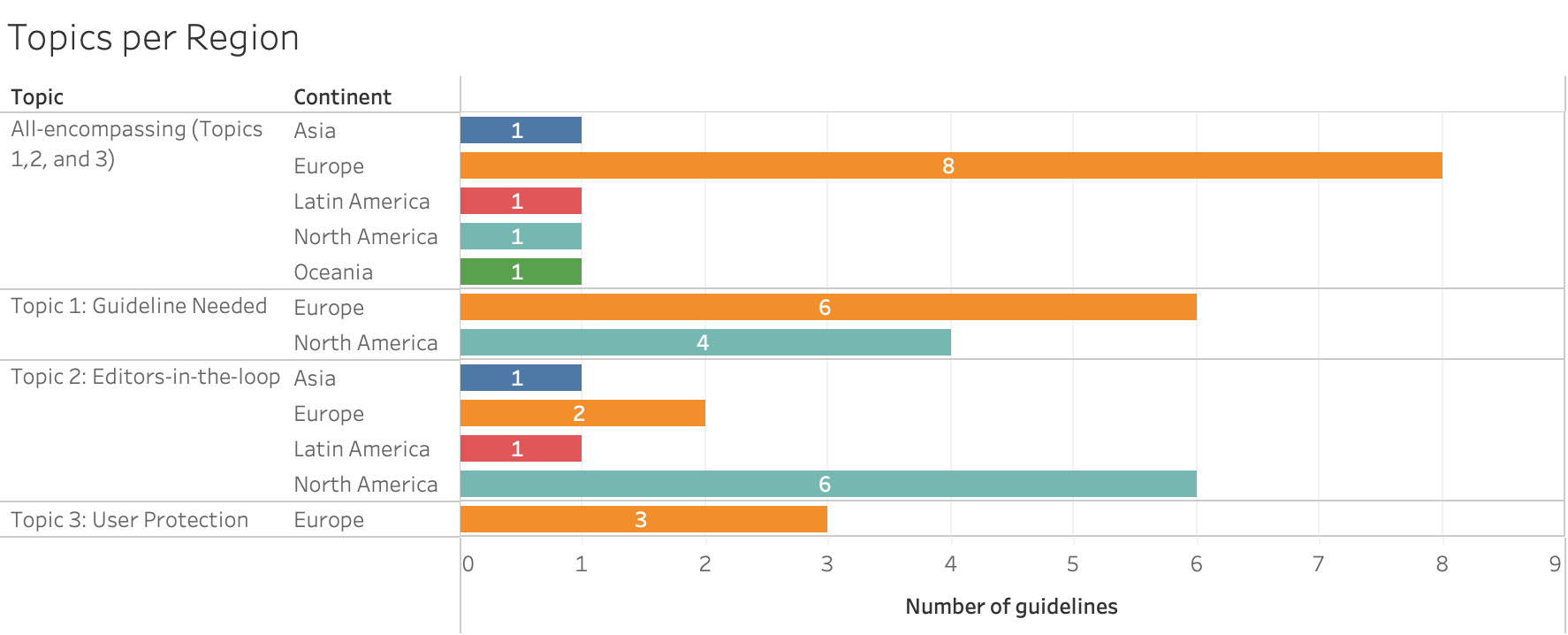}
\caption{AI Guidelines by Topics}
\label{fig:Fig3}
\end{figure}

To gain a better understanding of the main topics covered in each region, we created Figure \ref{fig:Fig3}, which illustrates the prevalent types of guidelines in each continent. As observed, all-encompassing guidelines (i.e., documents that cover all topics) (documents that cover all topics) are utilized in all continents. European media organizations are the most prolific in this regard, leading in frequency with eight guidelines. They also cover a wider range of area, as these organizations have issued AI guidelines across each distinct topics, particularly distinguishing themselves as the only continent that has produced user protection-focused guidelines. In contrast, North American media organizations are more focused on upholding journalistic values and curating guidelines for the responsible development of AI in newsrooms. In terms of absolute counts, European and North American media organizations are unsurprisingly at the forefront of institutionalization of AI deployment and use in the media industry. However, we do observe that the topic of "human-in-the-loop" is the only topic-specific guideline adopted by non-WEIRD countries, while most of them have issued all-encompassing guidelines.

\subsubsection{Topic 1: Need for Guidelines to Ensure Ethical and Responsible AI Development}

Building upon the efforts (Topic 2) already and will be made to protect the fundamental values and rights (Topic 3), these guidelines express a general direction of the responsible design, development, and use of AI in newsrooms. As elucidated from the aforementioned topics, both the promises and pitfalls of the AI-related solutions are clearly visible to the newsrooms, and hence these newsrooms pledge to create a programme that ensures a responsible and ethical development of AI-powered “services” and “systems” in newsrooms. Logically, guidelines are the first step towards such a programme; thus, the keyword “guideline” also prevails in this topic. 

The responsible development of AI should also align with the core "values" of newsrooms, and one solution could involve deploying specialized teams to enhance the performance of AI solutions, particularly in reducing bias. Several guidelines touch upon this topic, including those of German Public Service Media (PSM) Bayerischer Rundfunk (BR), the news agency Thomson Reuters, and two Dutch legacy media outlets, MediaHuis and Volkskrant, among others.

For instance, Thomson Reuters states, "Thomson Reuters will adopt the following Data and AI Ethics Principles to promote trustworthiness in our continuous design, development, and deployment of artificial intelligence ('AI') and our use of data." This indicates that Reuters' guideline emphasizes the responsible use of AI systems in their news production processes. Responsible AI practices, as outlined by Reuters, prioritize privacy and security throughout the design and development of AI products and services while also ensuring that AI products and services are understandable. Similarly, Bayerischer Rundfunk (BR) highlighted its role as public service media, emphasizing that the technologies used by newsrooms are ultimately accountable to the public and should be "public service minded."

\subsubsection{Topic 2: Editorial values in place}

Topic 2 reveals the need for editorial values is the most prominent theme. It reverberates with the concept of keeping “human-in-the-loop,” but more so on an editorial level to ensure a safe gatekeeping process throughout news production when AI is involved in ways illustrated in Topic 1. The keywords of this topic are, for example, “journalism,” “news,” “fact,” “research,” and “newsroom.” Organizations that principally cover this topic include for example The Guardian, leading American digital media WIRED, CNET and Business Insider. For example, Hong Kong Free Press’s guideline repeatedly refers to the need for humans in the loop, as evinced by their short line “owing to the technology’s aforementioned limitations [biases], we will confirm and double-check AI-generated output”. Similarly, Canadian legacy media also explicitly stated that “No CBC journalism will be published or broadcast without direct human involvement and oversight. We will never put to air or online content that has not been vetted or vouched for a CBC journalist […] (an array of generative AI tools requires) a heightened level of skepticism and verification in our journalism.”

\subsubsection{Topic 3: Users’ Protection }

While Topic 2 emphasizes the importance of human-in-the-loop during the processes of news production, Topic 3 hints at the promises that these guidelines make to protect the users to avoid potential violation of individual rights during the encounters with AI on their respective news sites and mobile phone applications. This topic does not merely suggest that the risks of AI are well aware by the newsrooms, but it also demonstrates the core values that newsrooms always possess for their audience, regardless of AI. These values can be observed from the list of keywords, such as “trust”, “public”, “value” and “responsible”. Intriguingly, the guidelines that preeminently addresses the protection of users are PSM and regulators from the United Kingdom, both mentioning the actions that should be taken to protect the creators, publishers, and users of AI-generated content. Delving deeper into the guidelines, one can observe that these recommendations are action-oriented but not always concrete. For example, BBC’s Generative AI at the BBC suggests mitigating the challenges it may create, such as “trust in media, protection of copyright and content discovery”.  However, although the guideline signals the importance of aforementioned values while developing generative AI, it only vaguely describes that the BBC will “work with the tech industry, media partners and regulators to champion safety and transparency in the development of Gen AI and protection against social harms” without further elaborating on how to do so.

\subsection{Findings derived from in-depth close reading}

The findings from these diverse sources emphasize the dynamic nature of AI guidelines in the ever-evolving landscape of technology and ethics. These guidelines raise crucial concerns regarding the integration of AI in media. One primary concern is the unchecked algorithmic news generation, where AI, especially generative AI, can introduce falsehoods, errors, and even hallucinated content into news articles, potentially eroding journalistic credibility. There is also growing recognition of a widening skill gap needed to manage this evolving field. Another significant concern is the potential for workflow disruptions, as AI tools impact existing newsroom processes and challenge how human journalists and AI systems collaborate effectively. Misuse of AI to intentionally spread misinformation and disinformation is a pressing issue, facilitated by AI's ability to create realistic and authentic-looking altered content. Transparency and disclosure are emphasized by some organizations, highlighting the need to attribute AI-generated content to maintain trust and avoid ambiguity.

In sum, while AI offers promising opportunities in media, these findings underscore the necessity for careful consideration and ethical guidelines to balance the benefits of automation with the preservation of journalistic integrity and the mitigation of risks, such as false or misleading AI-generated content \cite{Attard2023GenJournalism}. The common thread among these organizations is a commitment to adapt and update their guidelines in response to the rapid progression of AI technology. These documents recognize that the policies and principles guiding AI use are not static; instead, they are living documents subject to continuous refinement and adjustment. This flexibility ensures these guidelines remain relevant, responsible, and aligned with evolving ethical considerations, industry best practices, and technological advancements. It reflects a commitment to staying at the forefront of AI ethics and maintaining accountability in an ever-changing digital landscape. 

Next, we delve into the key topics that have emerged in these guidelines, providing insights into how the media industry perceives the future of AI within its ecosystem.

\subsubsection{Removing Bias and Errors While Chasing Accuracy and Correctness by the Human Operator}

In the guidelines, bias, and errors in news organizations’ AI technology use are constantly highlighted. These media organizations underscore the critical importance of vigilance and caution when dealing with AI-generated content. They acknowledge that AI, like humans, is susceptible to errors and biases. This recognition prompts a cautious approach to AI utilization, emphasizing the need for human oversight and verification. For example, the Allgemeiner Nachrichtendienst (ANP) demonstrates the commitment to quality journalism by not solely relying on AI-generated content, emphasizing the importance of human verification in their processes. It strikes a balance between investigation and caution in its approach to AI-generated content, recognizing that “AI texts may look promising as much as they may contain errors (‘hallucinations’) and reaffirm biases (‘bias’).”

Bayerischer Rundfunk (BR) prioritizes creating bias-free content by striving toward dialect models in speech-to-text applications and algorithmic accountability. Its editorial checks remain a mandatory element, although they have evolved to encompass plausibility checks of causal structures in data and a rigorous examination of data source integrity. Its guideline states that “the check of every individual piece of content is replaced by a plausibility check of causal structures in the data.”

Similarly, Deutsche Presse-Agentur (DPA) strongly emphasizes the technical robustness and security of their AI systems. The organization highlights the need to “minimize the risk of errors and misuse” by ensuring that their AI technology is technically sound and reliable. As the Dutch organization, de Volkskrant, aptly puts it, “we are both investigative and cautious because AI texts may look promising as much as they may contain errors (‘hallucinations’) and reaffirm biases (‘bias’).” This acknowledgment highlights the nuanced nature of AI-generated content, which can exhibit both promise and pitfalls \cite{Attard2023GenJournalism}.

Mediahuis also stands vigilant against potential biases in AI systems, recognizing the risk of discrimination based on various factors. The Dutch conglomerate is acutely aware that “current AI tools may be prone to error and bias” and factor this into its approach to developing proprietary AI models and tools. Thus, Mediahuis adopts a comprehensive approach to data management and content quality. The conglomerate affirms its commitment to accuracy and fairness by stating, “[we] regularly check data for accuracy and biases, and securely store and delete it when required.” This commitment to data integrity underscores the importance of minimizing biases and ensuring accurate reporting through careful monitoring and storage practices.

The News Media Alliance shares a commitment to preventing unintended bias in AI-generated content, seeking to “avoid discriminatory outcomes” through diligent evaluation of their AI systems. The Radio Television Digital News Association (RTDNA) delves into the critical question of how newsrooms, stations, and parent companies navigate the use of AI. The association acknowledges the transformative power of AI programs in content modification. However, RTDNA cautions that AI’s capabilities may be limited in providing proper context, which can result in “facts misplaced or confusing content” without thoughtful human guidelines. It prompts users and developers to consider establishing guidelines for mitigating errors and bias associated with AI, emphasizing the need for clear and ethical standards.

In a similar vein, Wired highlights the susceptibility of current AI tools to both errors and bias, often resulting in uninspired and unoriginal writing. Wired also brings a practical perspective, acknowledging that despite the promise of AI, it is not infallible. The organization pragmatically asserts that “[in] practice, though, AI will make mistakes and miss things that a human would find relevant—perhaps so much so that it doesn’t save any time.” This underscores the importance of understanding AI’s limitations and the role of human oversight in ensuring accuracy and relevance in news content.

These findings reveal a multifaceted approach within news organizations, wherein they navigate the challenges of AI-generated content with a keen eye on bias and errors. They are striving to harness the potential of AI while minimizing its inherent risks. ANP’s emphasis on verification, BR’s quest for bias-free content, and DPA’s focus on technical robustness exemplify a commitment to quality journalism and ethical AI. Mediahuis and the News Media Alliance prioritize vigilance against bias, while RTDNA and Wired underscore the need for human oversight to contextualize AI-generated content effectively. They highlight the complex interplay of promise and risk in AI-generated content, emphasizing the vital role of human oversight, bias mitigation, and ongoing adaptation to ensure the integrity and accuracy of media content in the AI era.

\subsubsection{Legal, Privacy, Transparency, and Ethics: Responsible Principles}

Credibility, accountability, and truth are all underpinned by transparency in journalism \cite{Phillips2010TRANSPARENCYJOURNALISM}. When journalists are transparent, the public can better evaluate the trustworthiness of the news they consume. Similarly, transparency emerges as a recurring theme in these guidelines to address potential errors that these AI system might occur, with organizations committed to openly communicating when AI-generated creation commits mistakes. Organizations grapple with the challenges posed by AI-generated material, recognizing that it can introduce errors and inaccuracies in unpredictable ways. These concerns underscore the need for a nuanced approach to the ownership and reliability of AI-generated content.

Several organizations prioritize transparency by explicitly communicating the trustworthiness of AI systems. They employ content and user group restrictions, emphasizing the importance of data protection and privacy. Moreover, some organizations adhere to a “four-eyes principle” to ensure rigorous quality management and damage prevention, especially in critical applications such as AI face recognition. Transparency, as noted by de Volkskrant, extends to “explicitly report[ing] this [AI-generated content] and indicat[ing] where any errors are in the generated work.” Thus, the commitment to transparency serves as a cornerstone of accountability and trust-building with the audience.

Bias mitigation is another ethical concern, with several organizations expressing vigilance about potential biases in AI systems. They are steadfast in their commitment to ensuring that AI-generated content does not discriminate against individuals or groups based on various characteristics. This commitment aligns with the broader goal of promoting fairness and ethical content production.

Ownership, plagiarism, and intellectual property rights also feature prominently in the findings. Many esteemed media organizations emphasize the importance of safeguarding privacy and data integrity when dealing with AI-generated content. Organizations such as Adobe explicitly prohibit using generative AI features to produce content infringing on third-party rights. Respecting intellectual property and adhering to copyright laws is paramount to maintaining the integrity of media and creative content. For example, Mediahuis “respect[s] copyright, particularly when AI elaborates on or imitates a recognizable style, content, approach or imagery from human creators.”

These guidelines also shed light on careful handling of confidential sources and proprietary information. One prevalent theme across these organizations is the need for stringent confidentiality measures. These organizations express a strong commitment to avoiding privacy risks and respecting users’ privacy. This includes minimizing the collection and use of personal data, obtaining user consent where required, and informing users about the purposes of data collection for AI services. According to some guidelines, the principle of informed consent is crucial in maintaining ethical data practices. Organizations also emphasize the prohibition of entering confidential information, trade secrets, or personal data into AI tools. This practice aligns with a commitment to upholding the trust of journalistic sources and protecting sensitive information.

As stated by Associated Press (AP), staff is urged “to not put confidential or sensitive information into AI tools.” This cautionary approach reflects the recognition that AI, like any technology, must be handled with discretion to prevent potential privacy and data security breaches. Regular documentation, monitoring, and review are integral to ensuring that data is handled securely and that AI algorithms serve diverse audiences equally and fairly. These findings underscore the significance of ongoing monitoring, documentation, and review to maintain data security and algorithmic fairness. 

In addition to data privacy, ethical considerations extend to ensuring that AI applications align with legal provisions and fundamental rights, including human dignity, freedom, equality, and non-discrimination. These principles form the bedrock upon which AI systems should operate. Compliance with applicable laws, such as data protection regulations, is crucial to protect personal data from unauthorized access and breaches. Furthermore, there is a strong emphasis on safeguarding publishers’ intellectual property (IP) and brands.

Most of these guidelines highlight the industry’s commitment to transparency, ensuring that audiences are fully informed when AI is used in media content creation. The Norwegian media outlet Verdens Gang (VG) states: “AI-generated content must be labeled clearly, and in a way that is understandable.” Similarly, The Guardian reinstates “[w]e will be open with our readers when we do this.” Such transparency builds trust and credibility in AI-augmented journalism, according to these organizations.

\subsubsection{Being Accountable and Respecting Journalistic Values}

These guidelines see the integration of AI into newsrooms with a dual focus on accountability and preserving journalistic value. The topic of accountability in journalism in the age of AI is of paramount importance to maintain trust and uphold journalistic integrity. Aftonbladet emphasizes its responsibility for all published content, including that produced with AI. The Czech news outlet, CTK, strongly emphasizes accountability by stating, “[d]eployers of GAI systems should be held accountable for system outputs.” Reuters underscores the importance of accountability and data ethics, stating, the organization “will implement and maintain appropriate accountability measures for our use of data and our AI products and services.” This is the only publisher whose commitment to accountability extends to data and AI applications.

The USA Today Network highlights the accountability and responsibility of mews practitioners when using AI-generated content by reinforcing that these professionals “must take responsibility for any errors or inaccuracies in the AI-generated content they use. They must be accountable to their audience and take corrective action if errors are found.” This guidance emphasizes the role of journalists in ensuring the quality and accuracy of AI-generated content.

Similarly, VG emphasizes that AI-generated content should be treated like other sources of information. For the organization, “journalists must treat the use of content created with the help of AI (generative AI) in the same way as other sources of information; with open curiosity and with caution.” This guidance ensures that AI-generated content undergoes the same scrutiny as journalistic content from traditional sources. The Canadian Globe and Mail is more contentious by stating that “AI tools like ChatGPT should not be used to condense, summarize or produce writing for publication. Doing so would potentially risk our reputation and the confidentiality of our reporting”.

The guidelines shed light on the enduring value of journalism in the context of AI integration. The Brazilian news outlet Nucleo makes it clear that AI should be applied to facilitate journalism rather than replace it. This aligns with the idea that AI should complement and enhance journalistic work rather than supplant it. AP emphasizes the core values of accuracy, fairness, and speed in journalism, stating that AI can serve these values and enhance the way they work. Equally, the US media organization makes it clear that AI is not seen as a replacement for journalists, underlining the indispensable role of human reporters.

This association to journalism values are also mentioned in other forms. The Guardian acknowledges its commitment to adopting emerging technologies while staying true to its mission of “serious reporting that uncovers facts, holds the powerful accountable, and interrogates ideas and arguments,” reflecting a balance between technological innovation and preserving the essence of quality journalism. The Austria Press Agency (APA) defends that “[all services and products of the APA Group, including AI-based solutions, are based on these high editorial quality standards when applicable,” driving their AI guidelines by editorial values. CBC/Radio-Canada also sees that the heart of the AI approach are “the principles of trust, transparency, accuracy and authenticity that are already core to our Journalistic Standards and Practices (JSP).” This process institutionalizes AI guidelines within the Journalistic Code of Principles, aiming to standardize and establish uniformity in these courses of action. This effort seeks to make the actions of their employees or divisions more predictable and, presumably, of higher quality, which does not include cultural and social factors \cite{Holder-Webb2012TheEthics}.

\subsubsection{No Replacement, Always Human-In-The-Loop}

The recurring theme across all guidelines regarding human-in-the-loop is the clear recognition of the irreplaceable role of humans in the media process, even when incorporating AI technologies. For example, Aftonbladet underscores that all published material has to be reviewed by a human and falls under their publishing authority, emphasizing the final human oversight in their editorial process. ANP maintains the “human-machine-human” approach, indicating that the chain of thinking and decision-making in journalism begins and ends with humans, with AI serving as a supporting tool. CBC reassures its audience that no CBC journalism will be published without direct human involvement and oversight. The Canadian organization is committed to fact-based, accurate, and original journalism done by humans for humans. Similarly, Stuff insists on human oversight and fact-checking for AI-generated content, holding it to the same high standards as content written by professional journalists. Wired emphasizes that the judgment required for journalism depends on understanding both the subject and readership, something AI cannot fully achieve. The publisher also highlights the importance of human evaluation and discernment in choosing which AI-generated suggestions to pursue.

In summary, these guidelines reflect a strong commitment to maintaining the essential role of humans in journalism, with AI serving as a tool to enhance efficiency and support various tasks but not as a replacement for the unique qualities, judgment, and creativity that human journalists bring to the field \cite{Munoriyarwa2023ArtificialNewsrooms}. The consensus among these media organizations is that AI should augment, facilitate, and support journalism while always being subject to human supervision and final editorial judgment. This approach is seen as a way to ensure journalistic integrity, accuracy, and relevance in a rapidly evolving media landscape. In other words, preserve the boundaries of journalistic work \cite{2015BoundariesJournalism}.

\subsubsection{Managing Top-Down Collaborations}

The successful adoption of AI technologies in newsrooms hinges on effective collaboration between journalists and AI specialists. As stated by the AP, “[j]ournalists who work well with data scientists and computational journalists are best positioned to thrive in an AI-assisted newsroom.” This highlights the importance of a seamless partnership between individuals with journalistic skills and those with AI and data analysis expertise. This collaborative synergy ensures that AI complements and augments journalistic efforts rather than supplanting them. Additionally, the BR emphasizes collaboration with startups and universities to leverage AI competencies in the region. This underscores the value of engaging with external partners to harness AI’s full potential, showcasing the importance of teamwork and knowledge-sharing beyond the newsroom.  However, there is a growing concern about this trend. Researchers point out the growing power of big tech companies, such as Amazon, Microsoft, and Google, in shaping how AI is being developed and used on a large scale \cite{de-Lima-Santos2021ArtificialOutlook,vanderVlist2024BigIntelligence}.

Top-down guidance is highly mentioned in these guidelines. To maintain ethical standards and responsible AI usage, editorial leadership plays a pivotal role. In Australia, a study has indicated that Generative AI is being deployed in media outlets using a centralized approach (top-down), while other organizations encourage a more bottom-up process where teams have the freedom to test ideas and report their results.\cite{Attard2023GenJournalism} The CBC’s commitment not to use AI-powered identification tools for investigative journalism without proper permissions is a clear example of top-down guidance. Furthermore, The Guardian insists on senior editor approval for any significant AI-generated content, ensuring human oversight and specific benefits. Similarly, Volkskrant enforces a strict top-down approach, requiring permission from chiefs and editors-in-chief for the use of generative AI. This hierarchical approach ensures that AI applications align with journalistic integrity and editorial standards. These guidelines might not reflect real-world practices, as a recent Australian study revealed \cite{Attard2023GenJournalism}. It might have a gap between policy and action, as media organizations are adopting a "test and learn" approach to Generative AI. This can take two forms: top-down decisions by managers or bottom-up experimentation encouraged by leadership.

In summary, integrating AI into media thrives on a collaborative spirit between journalists and AI experts. Effective teamwork ensures AI enhances journalistic capabilities. Moreover, top-down guidance and ethical standards are seen as essential for responsible AI use. 

\section{Discussion and conclusions}

The findings presented in this study reveal a comprehensive view of how media organizations are navigating the integration of artificial intelligence (AI) in journalism while upholding ethical principles and journalistic values. These insights are crucial in the context of digital inequalities, isomorphism, and institutional theory, as they shed light on the dynamic nature of AI guidelines and their evolving role in media organizations worldwide.

These guidelines address several critical concerns associated with the integration of AI into media, including issues related to algorithmic news generation, skill gaps, workflow disruptions, misuse of AI for disinformation, transparency, and disclosure. All these organizations share a focus on keeping their guidelines up-to-date as artificial intelligence advances quickly. This reflects a commitment among mostly WEIRD media organizations to adapt and update their guidelines to keep pace with the rapidly evolving landscape of AI technology. This adaptability ensures that these guidelines remain relevant, responsible, and aligned with the ever-changing ethical considerations, industry best practices, and technological advancements in their contexts. Nevertheless, it is important to note that many of these guidelines lack comprehensive details on how to address specific ethical principles. Only a limited number of guidelines offer basic and straightforward examples.

Furthermore, these guidelines lack a specific characterization of the uses of generative artificial intelligence in these newsrooms. Consequently, they fail to specify which programs can actually be used by practitioners. This lack of specificity particularly impacts the description of ethical guidelines, which become more generic without a clear description of AI applications. In other words, these guidelines fail to explain and describe the current or potential uses of these technologies at different stages of the news value chain.

The literature on digital inequalities, isomorphism, and institutional theory provides a useful framework for understanding the dynamics at play in these guidelines. Digital inequalities are brought to the forefront by the absence of guidelines, which in turn widens the skill gap necessary for effectively managing AI in media \cite{Jamil2023EvolvingPakistan}. Viewed through the lens of Institutional Theory, which emphasizes the replication of established norms and practices \cite{Gondwe2023CHATGPTAI,Jamil2023EvolvingPakistan}, this pattern is evident in the reviewed guidelines. 

Our convergent methodology reveals that these guidelines exhibit institutional isomorphism, leading to convergence and ultimately homogeneity over time \cite{Holder-Webb2012TheEthics}. This phenomenon can be attributed to the rapid emergence of generative AI. In a neo-institutional model, alignment with the institutional environment is crucial for organizational survival. These coercive pressures may have contributed to the similarities observed in these guidelines. The normative professionalization of journalism has influenced these guidelines to adhere to journalistic norms and values. Many media organizations appear to be extending their existing journalistic values to the AI era rather than fundamentally adapting them to the diverse and evolving media landscapes of their regions. Lastly, the mimetic elements in these guidelines reflect the characteristics of journalistic innovation, which tend to be incremental, involving the emulation or modification of certain components or designs and making small improvements in news products \cite{Paulussen2016InnovationNewsroom}.

These isomorphism properties can further lead to institutional pressures or mechanisms of influence \cite{Vos2019TheorizingTheory}. While these guidelines cover critical issues related to transparency, bias mitigation, and human oversight, they may not fully account for the specific risks and challenges \cite{Munoriyarwa2023ArtificialNewsrooms} tied to the intricacies of language, ethnicity, and other contextual factors prevalent in the Global South \cite{Mabweazara2021TowardsJournalists}.

In the comprehensive analysis of AI guidelines within media organizations, several key observations have come to the fore, shedding light on the challenges and opportunities presented by AI in the media landscape. Notably, these guidelines predominantly emanate from early-adopter regions, often WEIRD countries and Anglo-Saxon nations. This observation aligns with prior research, emphasizing that AI tools are mainly deployed using data and models from these regions, particularly those using the English language \cite{Brodie2023DataPlanet}. 

For example, the concept of “human-in-the-loop” is central to these guidelines, highlighting the importance of maintaining a boundary between machine and human intelligence \cite{Motoki2023MoreBias}\cite{Feng2023FromModels}. While generative AI can automate certain tasks, such as data analysis or content production, it cannot replicate the nuanced judgment and critical thinking skills of human journalists. The reliance on AI could result in the loss of journalistic integrity and diversity of perspectives in news reporting. However, it is crucial to acknowledge that while some issues have been identified and addressed in the guidelines, there remain recognizable risks associated with AI technologies in the Global South, particularly related to color, ethnicity, and other socio-cultural nuances \cite{Brodie2023DataPlanet}. These guidelines do not always offer comprehensive solutions for potential problems, leaving room for further exploration of these issues in a regional context. 

The current discussion on ethical guidelines acknowledges important principles but lacks a crucial element: how these principles translate into action for news companies that rely on third-party AI applications and models, such as GPT, Bart, Midjourney, among others. Many of these ethical values may not be directly controllable by the media company itself, raising a vital concern on the use of these tools and models. Without a clear explanation of how news organizations will ensure adherence to these principles within third-party AI tools/models, the guidelines risk becoming aspirational rather than operational. In their current form, these guidelines may primarily serve as a public statement of the company's beliefs, rather than a practical guide for media practitioners working with AI-powered technologies in these media outlets. To strengthen these guidelines, these organizations should explore and address this gap. They could outline concrete steps that media practitioners can take to promote ethical AI use within their workflows, even when relying on external technologies. This could involve strategies for vendor selection, contractual agreements with AI providers, or internal auditing processes. Additionally, the guidelines could be expanded to provide journalists with practical guidance on how to critically evaluate the outputs of third-party AI tools and ensure alignment with the company's ethical principles.

The paucity of guidelines from Global South media organizations, with some exceptions like Brazil and Philippines, reflects an increasing gap in addressing the realities and challenges of AI adoption in these regions. One noticeable gap is the significant absence of AI guidelines in certain regions such as Africa and the Middle East. The scarcity of guidelines from these regions highlights a clear digital divide in the media landscape, which needs to be addressed \cite{Holder-Webb2012TheEthics}. This imbalance underscores the need for more inclusive and region-specific AI guidelines that consider the unique socio-cultural and linguistic contexts of non-WEIRD nations. For example, the absence of region-specific guidance raises concerns about the readiness of media organizations in Africa and the Middle East to tackle the ethical considerations and best practices essential for the integration of AI into their media operations, especially in areas where data and infrastructure may be lacking \cite{De-Lima-Santos2023GoogleEast}. AI systems are only as good as the data they are trained on, and if this is biased or incomplete, these systems can lead to skewed or inaccurate results. This could perpetuate stereotypes, reinforce bias, and limit the diversity of voices in the media landscape.

Moreover, the absence of these guidelines underscores the urgency of fostering an inclusive and global discourse on AI ethics, one that encompasses the perspectives and experiences of media organizations operating in regions with distinct linguistic, cultural, and sociopolitical dynamics. As Gondwe wisely puts, "[i]t is unrealistic to expect a single technology designer to create a universally applicable tool, as individual designers are shaped by cultural perspectives" \cite{Gondwe2024ArtificialFramework}. As AI continues to shape the media landscape globally, it is imperative that Global South countries are also equipped with comprehensive guidelines that address their unique needs, ensuring that ethical and responsible AI integration extends beyond the boundaries of WEIRD countries.

Importantly, the media industry is increasingly under pressure to adopt AI technologies across the news value chain. While AI has the potential to streamline workflows, enhance content creation, and improve audience engagement, there is a danger of over-reliance on AI, leading to a reduction in human oversight and decision-making. The reliance on AI could result in the loss of journalistic integrity and diversity of perspectives in news reporting. Another key issue with the push for AI adoption is the assumption that all media organizations need these tools. In reality, the use of AI should be driven by specific needs and goals, rather than a one-size-fits-all approach. For many smaller or niche media outlets, the costs associated with implementing AI systems may outweigh the potential benefits. Additionally, the complexity of AI technologies can present challenges for organizations with limited technical expertise or resources.

To effectively tackle the unique challenges, opportunities, and contextual nuances associated with AI adoption in these diverse and dynamic regions, it is imperative to expand our perspective on AI deployment beyond the well-resourced media organizations in the Western world. Media organizations should assess their specific needs and capabilities before investing in AI technologies, and ensure that human oversight and ethical considerations remain paramount in the use of these tools. Additionally, AI literacy plays a crucial role in how these technologies are perceived and utilized in these areas \cite{Munoriyarwa2023ArtificialNewsrooms}. Prior research has highlighted that AI skepticism in Africa is driven by several factors related to the lack of knowledge, including concerns about potential job losses, the associated costs of deploying AI, inadequate training, ethical dilemmas related to these emerging technologies, and doubts about their effectiveness in the democratic process \cite{Munoriyarwa2023ArtificialNewsrooms}. The absence of such information may impede AI development in the media ecosystem of the Global South. 

While this study provides valuable insights, it has certain limitations. First, it employs Latent Dirichlet Allocation (LDA) for analysis, which may not capture the fine-grained details that word-level embeddings or transfer learning-based models could offer. This limitation suggests that future research could explore alternative natural language processing techniques to enhance the analysis depth. Second, the issue of translation arises when considering guidelines from non-English-speaking regions, potentially introducing subtle nuances that might be lost in translation. This limitation underscores the importance of conducting further research with a focus on guidelines published in their native languages to ensure an accurate understanding. Finally, as a conclusion and future direction, it is essential to emphasize that while these guidelines serve as valuable tools, they are not one-size-fits-all solutions. This study advocates for the acknowledgment of local limitations and contextual differences when applying these guidelines in regions like the Global South, to ensure their ethical and practical relevance in diverse settings.

This research contributes to the understanding of AI guideline adoption within media organizations by providing a comprehensive analysis of existing guidelines from a global perspective. Our examination of the guidelines’ content reveals common themes related to transparency, bias mitigation, human oversight, and accountability, shedding light on shared ethical concerns across media organizations. It also offers insights into the geographical distribution of these guidelines, highlighting the dominance of Western nations, particularly North America and Europe. As AI continues to evolve, these guidelines should remain dynamic, adapting to emerging challenges and ethical considerations in the ever-changing digital landscape. The adoption of institutional best practices from WEIRD countries, tailored to the specific needs and challenges of the Global South, can play a pivotal role in narrowing the digital divide and fostering a more inclusive digital ecosystem worldwide. However, the replication of Western media guidelines is not a solution to these problems. Bridging the digital divide is not just a matter of providing access; it is a prerequisite for achieving sustainable development goals and improving overall quality of life. Recognizing the complexity of this issue and the importance of addressing it is the first step toward creating a more inclusive and equitable digital future for the media industry in the Global South.

\section*{Data Availability }
Data are publicly available \href{https://osf.io/hc9vb/?view_only=a0719618ce9a4b41bcdf6e3404c365f5}{here} respecting the European General Data Protection Regulation. 

\section*{Competing interests}
The authors declare that they have no competing interests.

\section*{Author's contributions}
    MFLS: Mathias-Felipe de-Lima Santos.
    WNY: Wang Ngai Yeung.
    TD: Tomás Dodds. 
    Conceptualization: MFLS and TD; methodology: MFLS and WNY; validation: MFLS and WNY; formal analysis: MFLS and WNY; data  curation: MFLS and WNY; writing—original draft preparation: MFLS, WNY, and TD;  writing—review and editing: MFLS, WNY, and TD; visualization:  MFLS and WNY; supervision: MFLS; funding acquisition: MFLS. All authors have read and agreed to the published version of the manuscript.
    
\bibliography{library.bib}

\end{document}